\documentclass[aps,reprint,superscriptaddress]{revtex4-2}
\usepackage{graphicx}
\usepackage{booktabs}
\usepackage{multirow}
\usepackage{color}
\usepackage[normalem]{ulem}
\usepackage{amsmath,amssymb}
\usepackage{bm}

\begin{document}

\title{Ferromagnetic Order of Reduced Magnetic Moments in a Frustrated Sawtooth Chain of the Magnetic Semiconductor ZnYb$_2$S$_4$}

\author{Shinji Okada}
\affiliation{Department of Quantum Matter, Graduate School of Advanced Science and Engineering, Hiroshima University, Higashi-Hiroshima 739--8530, Japan}

\author{Hiroto Suzuki}
\email{Present address: Institute for Solid State Physics, the University of Tokyo, Kashiwa 277-8581, Japan}
\affiliation{Department of Quantum Matter, Graduate School of Advanced Science and Engineering, Hiroshima University, Higashi-Hiroshima 739--8530, Japan}

\author{Nonoka Higa}
\affiliation{Department of Quantum Matter, Graduate School of Advanced Science and Engineering, Hiroshima University, Higashi-Hiroshima 739--8530, Japan}

\author{Yasuyuki Shimura}
\affiliation{Department of Quantum Matter, Graduate School of Advanced Science and Engineering, Hiroshima University, Higashi-Hiroshima 739--8530, Japan}

\author{Takanori Taniguchi}
\affiliation{Institute for Materials Research, Tohoku University, Sendai 980-8577, Japan}

\author{Takahiro Onimaru}
\email{onimaru@hiroshima-u.ac.jp}
\affiliation{Department of Quantum Matter, Graduate School of Advanced Science and Engineering, Hiroshima University, Higashi-Hiroshima 739--8530, Japan}

\begin{abstract}
In a sawtooth spin chain, competing nearest- and next-nearest-neighbor interactions suppress long-range order, yielding novel quantum states such as a spin-dimer singlet, 1/2 magnetization plateau, and spin contraction.
Here, we investigate the magnetic properties of the orthorhombic semiconductor ZnYb$_2$S$_4$, in which Yb$^{3+}$ ions with an effective spin-1/2 form a sawtooth chain along the $b$-axis.
The specific heat exhibits a sharp peak at ${T}_{\rm m}$ $=$ 1.4 K, at which the magnetic entropy $S_{\rm m}$ reaches only 27\% of $R$ln2.
This reduced $S_{\rm m}$ at $T_{\rm m}$ indicates the entropy release of the ground state doublet of Yb$^{3+}$ even for $T$ $>$ $T_{\rm m}$.
The isothermal magnetization $M(B)$ at 0.28 K exhibits hysteresis for $\left|B\right| \leq 0.2$ T and increases monotonically for $B > 0.2$ T. 
The spontaneous magnetization is only 0.1 ${\it \mu}_{\rm B}$$/$Yb, an order of magnitude smaller than that expected for the ground state doublet of Yb$^{3+}$. 
Moreover, in powder neutron diffraction measurements, no superlattice reflections due to antiferromagnetic order are observed for $T$ $<$ $T_{\rm m}$.
Therefore, in the ground state, the Yb moments are ferromagnetically aligned, but their amplitude is reduced by magnetic frustration in the sawtooth Yb chain.
\end{abstract}

\date{\today}

\maketitle

\section{Introduction}

In correlated electronic systems, geometrical frustration is of great interest since competing magnetic exchange interactions suppress long-range order and give rise to novel quantum states, including spin-dimer singlets, magnetization plateaus, and quantum spin liquids \cite{Greedan01,Balents10,Savary17,Shimizu03}.  
These frustration effects have been investigated not only in $ d$-electron spin systems but also in $f$-electron systems with strong spin--orbit coupling.
In the pyrochlore compounds $R_2$Ti$_2$O$_7$ ($R$ $=$ Dy and Ho), the residual entropy of the spin-ice state and the ice-rule-breaking spin flips have been observed \cite{Harris97,Bramwell01,Sakakibara03}, and the emergence of magnetic monopoles has also been proposed \cite{Castelnovo08,Kadowaki09,Fennell09}.  
In Pr$_2$Ir$_2$O$_7$, the chiral spin liquid state could manifest itself, owing to the underlying quantum criticality \cite{Nakatsuji06,Machida10,Tokiwa14b}

Recently, Yb-based insulators and semiconductors with 4f$^{13}$ configuration featuring a triangular Yb lattice have attracted considerable interest, as strong spin--orbit coupling and crystalline electric field (CEF) effects give rise to effective spin-1/2 systems with highly anisotropic exchange interactions \cite{Li15a,Li15b,Li16,Baenitz18,Ranjith19a,Ranjith19b,Xing19,Yamamoto14}.
In YbMgGaO$_4$, a gapless quantum spin liquid state has been proposed, as the effective spin-1/2 moments show no phase transition down to 0.048 K \cite{Li15a,Li15b,Li16}. 
Among NaYb$X_2$ ($X$ $=$ O, S, and Se) \cite{Baenitz18,Ranjith19a,Ranjith19b} and CsYbSe$_2$ \cite{Xing19}, which lack long-range magnetic order in zero field, NaYbSe$_2$ and CsYbSe$_2$ exhibit 1/3 magnetization plateaus \cite{Yamamoto14}.
On the other hand, in the semiconductor YbCuS$_2$ with a Yb$^{3+}$ zigzag chain, a first-order phase transition appears at $T_{\rm o}$ $=$ 0.95 K~\cite{Ohmagari2020-2,Ohmagari2020}, below which an incommensurate alignment of reduced ordered Yb moments has been revealed \cite{Onimaru25,Hori2023}. 
In addition, the spin-lattice relaxation rate $T_{1}^{-1}$ shows a $T$-linear behavior, indicating a gapless spin excitation~\cite{Hori2023}. 
These observations agree with theoretical calculations based on the density matrix renormalization group with off-diagonal $\Gamma$-type anisotropic exchange interactions \cite{Saito2024,Saito2024-2,Saito2024-3}.

 \begin{figure}[b]
  \centering
       \includegraphics[width=85mm]{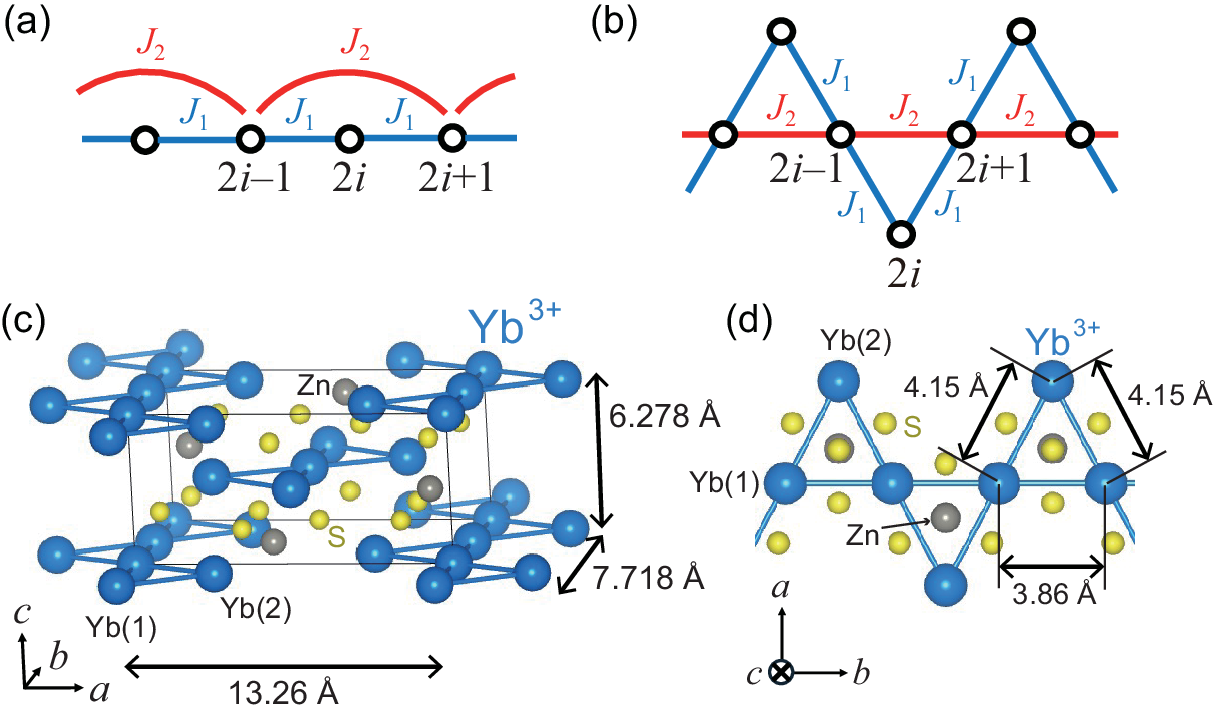}
 \caption{(Color online)
(a) One-dimensional spin chain with nearest- and next-nearest-neighbor exchange interactions, $J_{\rm 1}$ (blue) and $J_{\rm 2}$ (red).
Note that $J_{2}$ acts on every other site. 
(b) A sawtooth spin chain with $J_1$ and $J_2$ equivalent to the one-dimensional model. 
(c) Orthorhombic crystal structure of ZnYb$_2$S$_4$ with the space group of $Pnma$. There are two Yb$^{3+}$ sites, which are labeled Yb(1) at the 4$a$ site ($\bar{1}$) and Yb(2) at the 4$c$ site ($m$).
(d) The Yb$^{3+}$ ions of Yb(1) and Yb(2) form a sawtooth chain along the $b$-axis. 
 }
  \label{fig1}
\end{figure}

Another platform of a magnetically frustrated system is the sawtooth spin chain. 
As shown in Figs. \ref{fig1}(a) and \ref{fig1}(b), competition between the nearest-neighbor interaction $J_1$ and the next-nearest-neighbor interaction $J_2$ induces magnetic frustration. 
The Hamiltonian for the sawtooth spin chain can be written as
\begin{equation}
\begin{aligned}
\mathcal{H} =  J_{1} \sum_{i} &( \bm{S}_{2i-1} \cdot \bm{S}_{2i} + \bm{S}_{2i} \cdot \bm{S}_{2i+1}) \\ 
&+ J_{2} \sum_{i} \bm{S}_{2i-1} \cdot \bm{S}_{2i+1}
 \label{eq:sawtooth}
 \end{aligned}
 \end{equation}
, where $\bm{S}$ are spin operators and $J_{\rm 1}$, $J_{\rm 2}$ (\(> 0\)) are antiferromagnetic (AFM) exchange interactions. 
This model has been theoretically studied for decades~\cite{Kubo93,Penc96,Nakamura96,Sen96,Blundell03,Richter08,Jiang15,Metavitsiadis20,Richter22,Rausch25}. 
Its ground states are predicted as a function of $J_{\rm 2}/J_{\rm 1}$ as follows~\cite{Blundell03,Jiang15,Rausch25}: 
(1) When $J_{\rm 2}$ is much smaller than $J_{\rm 1}$, the $J_{\rm 1}$ term dominates, and the ground state is an AFM ground state. 
(2) For $0.5 < J_{\mathrm{2}}/J_{\mathrm{1}} < 1.5$, a gapped spin-dimer state is formed.
(3) For larger $J_{\rm 2}$, the ground state becomes a noncollinear magnetic order. 
In this phase, a 1$/$2 magnetization plateau may appear in the isothermal magnetization. 
Indeed, in Cu$_2$Cl(OH)$_3$, a possible 1$/$2 plateau has been observed \cite{Heinze21}.
On the other hand, the characteristic properties of 4$f$-electron sawtooth chain systems have not yet been reported.

In the present work, we focus on the rare-earth-based semiconducting olivine ZnYb$_2$S$_4$, which crystallizes in an orthorhombic structure with the space group $Pnma$ \cite{Lau06}.
As shown in Fig. \ref{fig1}(c) and \ref{fig1}(d), the Yb$^{3+}$ ions of Yb(1) at the 4$a$ site ($\bar{1}$) and Yb(2) at the 4$c$ site ($m$) form a sawtooth chain along the $b$-axis. 
The magnetic susceptibility follows the Curie--Weiss law between 50 and 350 K  \cite{Lau06}. 
The effective magnetic moment, ${\it \mu}_{\rm eff}$ = 4.72 ${\it \mu}_{\rm B}$/Yb, is close to that of a free Yb$^{3+}$ ion, 4.54 ${\it \mu}_{\rm B}$. 
The negative paramagnetic Curie temperature, ${\it \theta}_{\rm p}$ $=$ $-$75.2 K, indicates AFM interactions between the Yb moments. However, no phase transition has been reported above 1.8 K. 
Here, we synthesize polycrystalline ZnYb$_2$S$_4$ and investigate its magnetic properties to clarify whether a phase transition occurs at lower temperatures and how magnetic frustration in the sawtooth Yb chain affects the formation of the ground state.

\section{Experimental}

Polycrystalline samples of ZnYb$_2$S$_4$ were synthesized by a solid-state reaction method. 
Yb metal was reacted with S powder to synthesize Yb$_2$S$_3$ \cite{Schleid96,Chen19} in a sealed, evacuated quartz tube at 800$^\circ$C for 2 days.
The Yb$_2$S$_3$ and ZnS mixed pellet, sealed in an evacuated quartz tube, was heated to 1000$^\circ$C and held for 5 days.
The synthesized samples were characterized using powder X-ray diffraction and electron-probe microanalysis (EPMA).
Rietveld analysis of the powder X-ray diffraction pattern yielded lattice constants of $a$ $=$ 13.2561(2)~\AA, $b$ $=$ 7.71522(12)~\AA, and $c$ $=$ 6.26863(10)~\AA, 
which agree with those reported in the previous work \cite{Vollebregt82}.
EPMA confirmed that the atomic composition of the main phase is Zn : Yb : S = 1.02(1) : 2 : 3.80(3). 
In the X-ray diffraction pattern, small peaks from an impurity phase were observed; however, their intensities are smaller by 12\% than the most intense peak of the main phase. 
Using EPMA, we identified the impurity phase as Yb$_2$S, a compound that has not yet been reported.

Specific heat $C$($T$) was measured between 0.4 and 19 K in magnetic fields up to $B$~$=$~14~T using a Quantum Design PPMS and up to 300 K with a Cryogen Free Measurement System (CFMS-9T, Cryogenic Ltd.).
Magnetization $M$($T, B$) was measured from 1.8 to 300~K using a Quantum Design MPMS.
Measurements of $M$($T, B$) at low temperatures down to 0.28 K were conducted in magnetic fields of $B$ $\le$ 9 T using a laboratory-built capacitive Faraday magnetometer \cite{Sakakibara94} equipped in a $^3$He cryostat.

Powder neutron diffraction measurements were performed using the high-efficiency and resolution powder diffractometer (HERMES), installed at the JRR-3M reactor of Japan Atomic Energy Agency \cite{Ohoyama98,Nambu24}.
Neutrons with a wavelength of 2.1974(2)~{\AA} were obtained using the 331 Bragg reflection of a deformed single-crystal Ge monochromator. 
The powdered sample was sealed in a cylindrical vanadium capsule and cooled in a 1 K refrigerator down to 0.7 K.

\begin{figure}[b]
    \begin{center}
     \includegraphics[width=85mm]{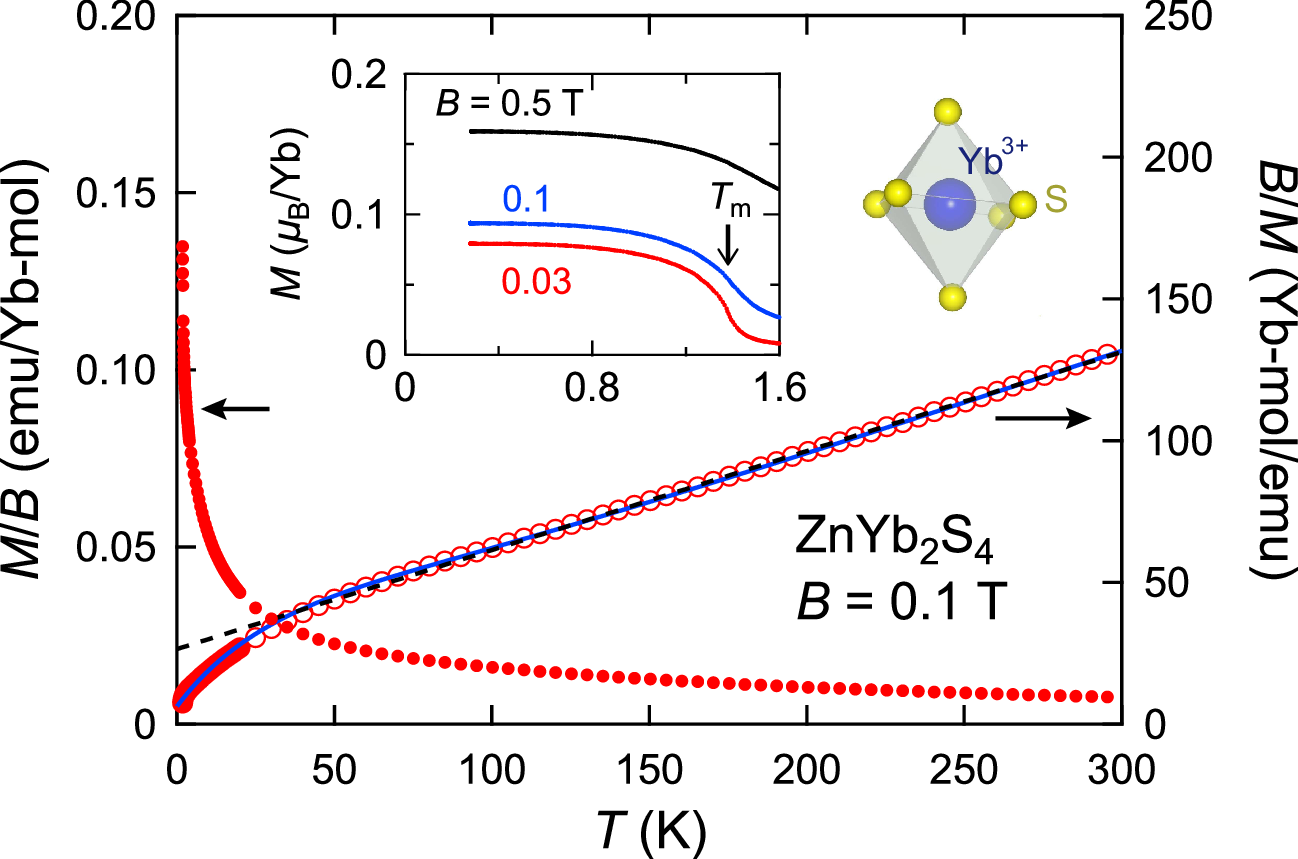}
     \caption{(Color online) Temperature dependence of the magnetic susceptibility $M(T)/B$ of ZnYb$_2$S$_4$ measured in $B$~$=$~0.1~T (left-hand scale), and the inverse susceptibility $B/M$ (right-hand scale). The broken line indicates a Curie--Weiss fit, and the solid (blue) line is a calculation considering the CEF effect for a Yb$^{3+}$ ion.
     The inset presents the temperature dependence of the magnetization in $B$ $=$ 0.03, 0.1, and 0.5 T. $T_{\rm m}$ denotes the magnetic ordering temperature. 
     The right inset shows the Yb(1) site encapsulated in a slightly distorted S octahedron.}
      \label{fig2}
    \end{center}
   \end{figure}

\section{Results and discussion}

\subsection{Magnetic susceptibility}

Figure~\ref{fig2} shows the temperature dependence of the magnetic susceptibility $M/B$ (left-hand scale) and its inverse $B/M$ (right-hand scale), measured at $B$ = 0.1 T. 
For $T$ $>$ 70 K, $B/M$ increases linearly with temperature, indicating the Curie--Weiss behavior. A Curie--Weiss fit to the $B/M$ data, shown with the broken line in Fig.~\ref{fig2}, yields an effective magnetic moment of ${\it \mu}_{\rm eff}$ = 4.78 ${\it \mu}_{\rm B}$ per Yb ion.
This value is close to the expected 4.54 ${\it \mu}_{\rm B}$ for a free Yb$^{3+}$ ion, implying that the magnetic properties are dominated by the Yb$^{3+}$ moments.
The Curie--Weiss fit gives a paramagnetic Curie temperature of ${\it \theta}_{\rm p}$ = $-$75.7 K.
The negative sign of ${\it \theta}_{\rm p}$ indicates predominant AFM interactions between the Yb moments.
These observations are consistent with the previous report~\cite{Lau06}.\\

To extract the CEF effect on the Yb$^{3+}$ ions, we analyze the temperature dependence of $B/M$.
ZnYb$_2$S$_4$ has two crystallographically distinct Yb sites, where both Yb ions are encapsulated similarly in slightly distorted octahedra of six S atoms, as shown in the inset of Fig. \ref{fig2} (Yb(1) site is shown). 
Thereby, we assume that the Yb ion is encapsulated in a regular octahedron, corresponding to the cubic point group.
The cubic CEF Hamiltonian for a Yb$^{3+}$ ion with $J$ $=$ 7/2 is
\begin{equation}
\mathcal{H}_{\mathrm{CEF}}=W\Bigl[\frac{x}{60}(O_4^0+5O_4^4)+\frac{1-\left|x\right|}{1260}(O_6^0-21O_6^4)\Bigr],
 \label{eq:CEF4}
 \end{equation}
 where $W$ and $x$ are CEF parameters, and $O_n^m$ are Stevens operator equivalents.
Introducing a mean-field parameter, $\lambda$, to account for exchange interactions among Yb moments, the inverse magnetic susceptibility $\chi^{-1}$ ($=$~$B/M$) is expressed as
\begin{equation}
\frac{1}{\chi} = \frac{1}{\chi_{\mathrm{CEF}}} - \lambda,
\end{equation}
where $\chi_{\mathrm{CEF}}$ is the CEF contribution to the magnetic susceptibility.
Fitting the $B/M$ data (the solid (blue) line in Fig.~\ref{fig2}) yields $W$ $=$ 25.7 K, $x$ $=$ $-$1.0, and $\lambda$ $=$ $-$6.19 Yb-mol$/$emu.
This $\lambda$ corresponds to a paramagnetic Curie temperature ${\it \theta}_{\rm p}$ = $-$17.7 K, indicating AFM interactions.
The cubic CEF splits the eight-fold $J$ multiplet into three levels: a $\Gamma_6$ doublet ground state, a first excited $\Gamma_8$ quartet at 309 K, and a $\Gamma_7$ doublet at 824 K. 
The well-isolated $\Gamma_6$ doublet ground state realizes an effective spin-1/2 system at low temperatures, with an expected value of the magnetic moment of 1.33 ${\it \mu}_{\rm B}$/Yb.

      \begin{figure}[t]
    \begin{center}
     \includegraphics[width=83mm]{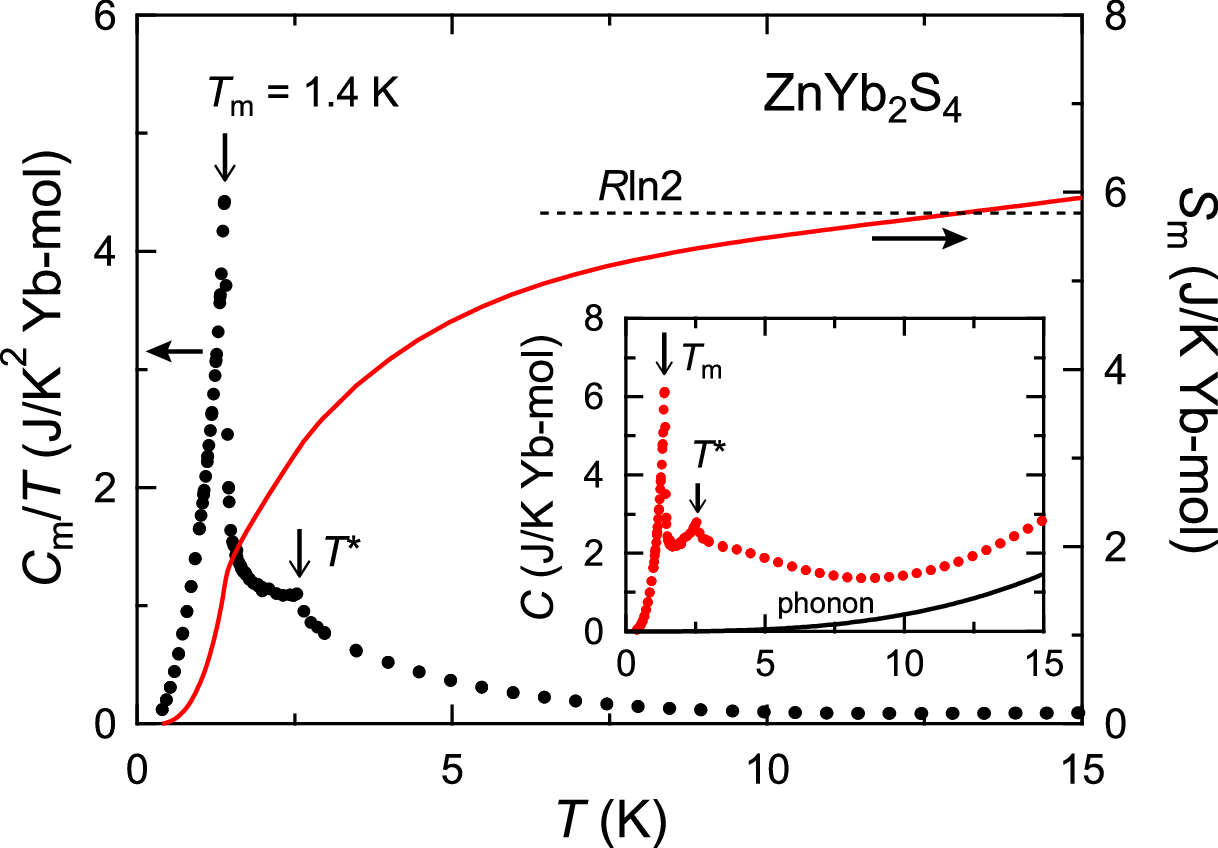}
     \caption{(Color online) Temperature dependence of the magnetic specific heat divided by temperature, $C_{\rm m}/T$ (left-hand scale), and the magnetic entropy, $S_{\rm m}$ (right-hand scale), of ZnYb$_2$S$_4$. $T_{\rm m}$ denotes the magnetic ordering temperature, and $T^{*}$ indicates an extrinsic anomaly due to an impurity phase. 
     The inset shows the temperature dependence of the specific heat, $C$($T$). The solid line represents the phonon contribution estimated using a Debye model. See the text in detail.
     }
      \label{fig3}
    \end{center}
   \end{figure}

   \begin{figure}[t]
    \begin{center}
     \includegraphics[width=77mm]{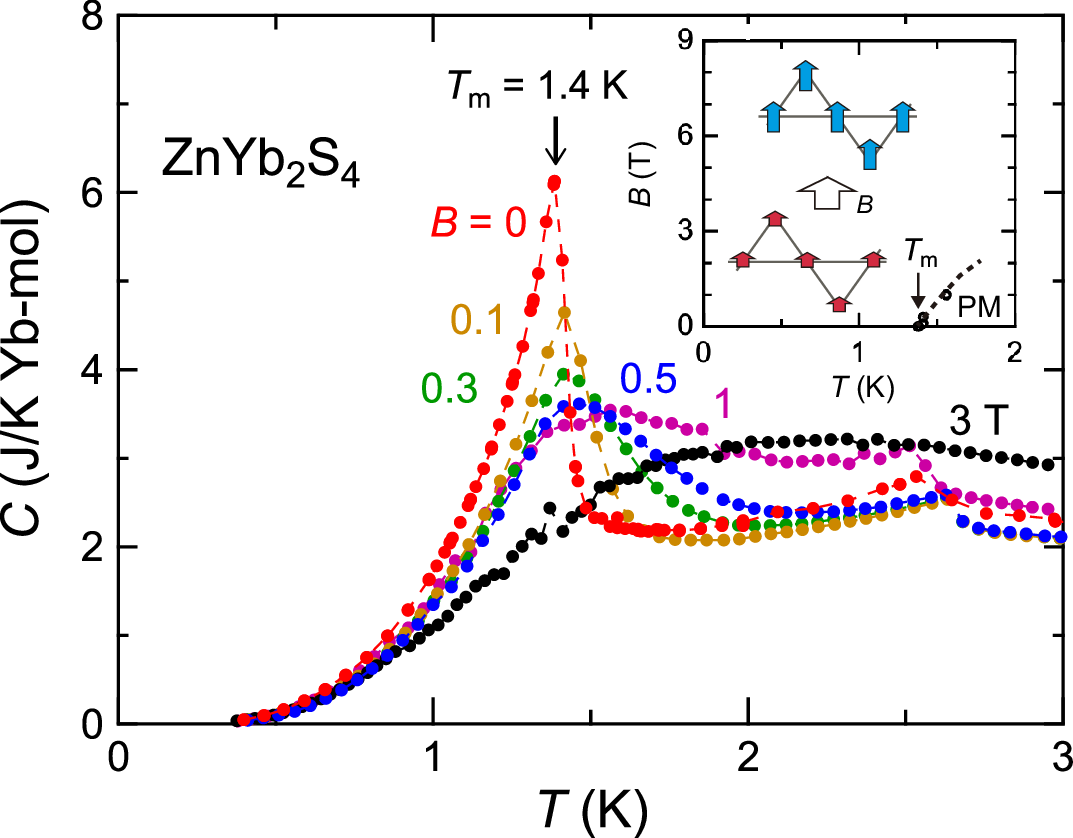}
     \caption{(Color online) Temperature dependence of the specific heat $C$ of ZnYb$_2$S$_4$ in $B$ $=$ 0, 0.1, 0.3, 0.5, 1, and 3 T.
     The inset shows the $B$--$T$ phase diagram. PM denotes the paramagnetic phase, and the broken line indicates the crossover.
     The illustration indicates that the reduced Yb moments (red arrows) on the sawtooth chain are developed (blue arrows) due to release of the magnetic frustration with the application of magnetic fields.}
      \label{fig4}
    \end{center}
   \end{figure}
   
\subsection{Specific heat}

The temperature dependence of the specific heat $C$ is presented in the inset of Fig.~\ref{fig3}.
A sharp peak appears at $T_{\rm m}$ $=$ 1.4~K, indicating a phase transition of the ground state doublet.
Another small peak is detected at $T^{*}$ $=$ 2.6 K, which is most likely due to an impurity phase, although its origin has not yet been clarified.
The main panel of Fig.~\ref{fig3} displays the temperature dependence of the magnetic specific heat divided by temperature, $C_{\rm m}$$/$$T$. 
$C_{\rm m}$ is obtained by subtracting the phonon contribution from the total specific heat $C$, as shown in the inset.
To evaluate this phonon contribution, we use the Debye model with a Debye temperature of $\theta_{\rm D}$ $=$ 250 K, which reasonably reproduces the $C(T)$ data between 30 and 200 K (not shown).
The magnetic entropy, $S_{\rm m}$($T$), plotted on the right-hand axis of Fig.~\ref{fig3}, is calculated by integrating the $C_{\rm m}$$/$$T$ data with respect to temperature.  
At ${T}_{\rm m}$, $S_{\rm m}$($T$) reaches 27\% of $R$ln2. 
This reduced entropy indicates that the entropy of the doublet ground state is released above ${T}_{\rm m}$, which is a hallmark of magnetic frustration.
As the temperature increases, $S_{\rm m}$($T$) reaches $R$ln2 at 13 K, indicating that the CEF ground state is a Kramers doublet.

Temperature dependence of $C$($T$) measured at various magnetic fields up to $B$ $=$ 3 T is shown in Fig. \ref{fig4}. 
At $B$~$=$~0.1 T, the peak broadens and shifts slightly to higher temperatures.
With a further increase in $B$, the peak is gradually suppressed, becomes broader, and shifts to higher temperatures.
Ultimately, a broad maximum appears at 2.3 K in $B$ = 3 T, which is moderately consistent with the Schottky specific heat expected from split energy levels.
This crossover behavior suggests that the ground state of ZnYb$_2$S$_4$ is ferromagnetic-like, and that the magnetic order is suppressed by an applied magnetic field.
In fact, as shown in the inset of Fig. \ref{fig2}, the magnetization drops sharply around $T_{\rm m}$ at $B$ $=$ 0.03 T as the temperature increases, whereas this change becomes indistinct at $B$ $=$ 0.5 T.
On the other hand, the peak of $C(T)$ at 2.6 K, which may originate from an impurity phase, disappears at $B$ $=$ 3 T.

\subsection{Isothermal magnetization}

\begin{figure}[t]
    \begin{center}
     \includegraphics[width=77mm]{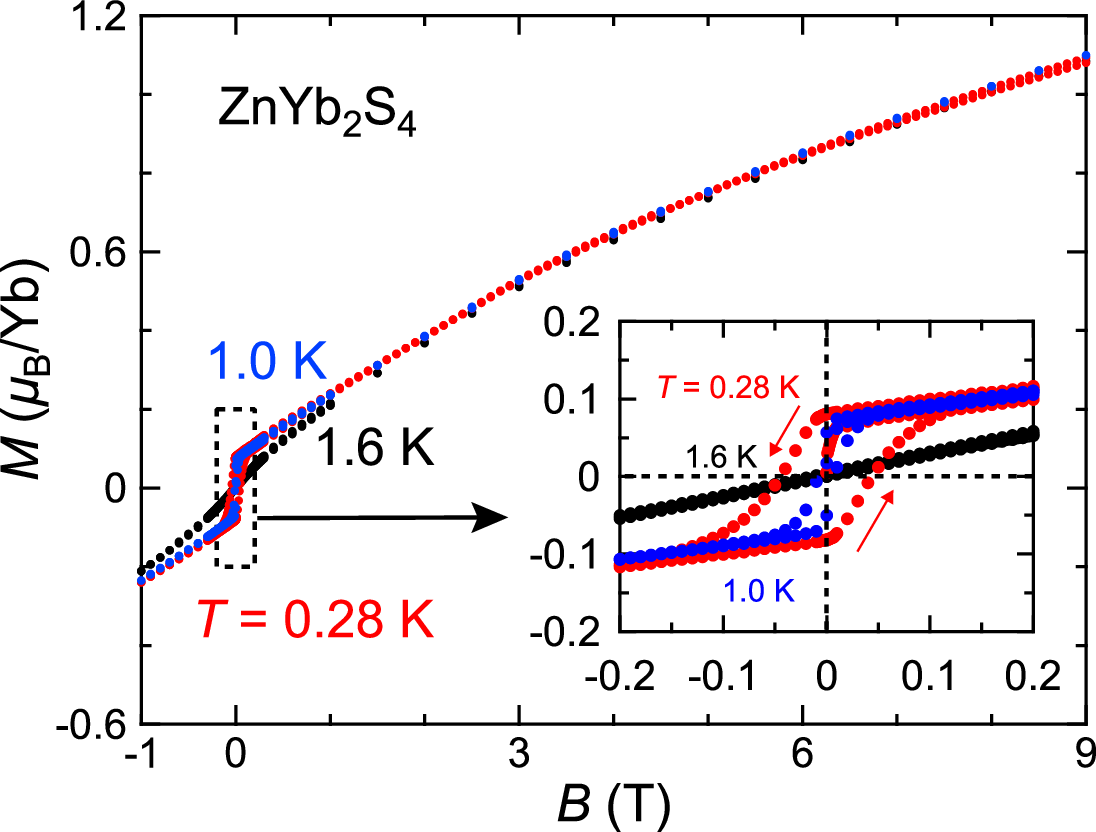}
     \caption{(Color online) Isothermal magnetization of ZnYb$_2$S$_4$ at $T$ = 0.28 (red), 1.0 (blue), and 1.6 K (black). The inset shows the data expanded for $|B|$ $\leq$ 0.2 T. 
     The arrows indicate the directions in which the magnetic fields change. }
      \label{fig5}
    \end{center}
   \end{figure}

Figure \ref{fig5} shows the magnetic-field dependence of the isothermal magnetization $M(B)$ at $T$ $=$ 0.28, 1.0, and 1.6~K. 
At $T$ $=$ 0.28 K ($<$ $T_{\rm m}$), $M(B)$ exhibits hysteresis for $\left|B\right| \leq 0.2$ T (see inset) and increases monotonically for $B$ $>$ 0.2 T, reaching 1.1 ${\it \mu}_{\rm B}$/Yb at 9 T.
Hysteresis is also present at 1.0 K, but disappears at 1.6 K, above $T_{\rm m}$.
The spontaneous magnetization at 0.28 K is about 0.1 ${\it \mu}_{\rm B}$/Yb, roughly an order of magnitude smaller than the expected 1.33 $\mu_{\rm B}$/Yb for the ground state doublet of Yb$^{3+}$. 
We also find that $M(B)$ for $B$ $\ge$ 1 T is almost identical between 0.28 K and 1.6 K, i.e., both below and above $T_{\rm m}$.
This suggests that the Yb moments may be recovered by applying magnetic fields, which release the magnetic frustration.
It is noted that 1/2 magnetization plateau does not appear for $T$ $<$ $T_{\rm m}$, contrary to the theoretical prediction for the AFM sawtooth chain~\cite{Rausch25}.

  \begin{figure*}[t]
    \centering
    \includegraphics[width=160mm]{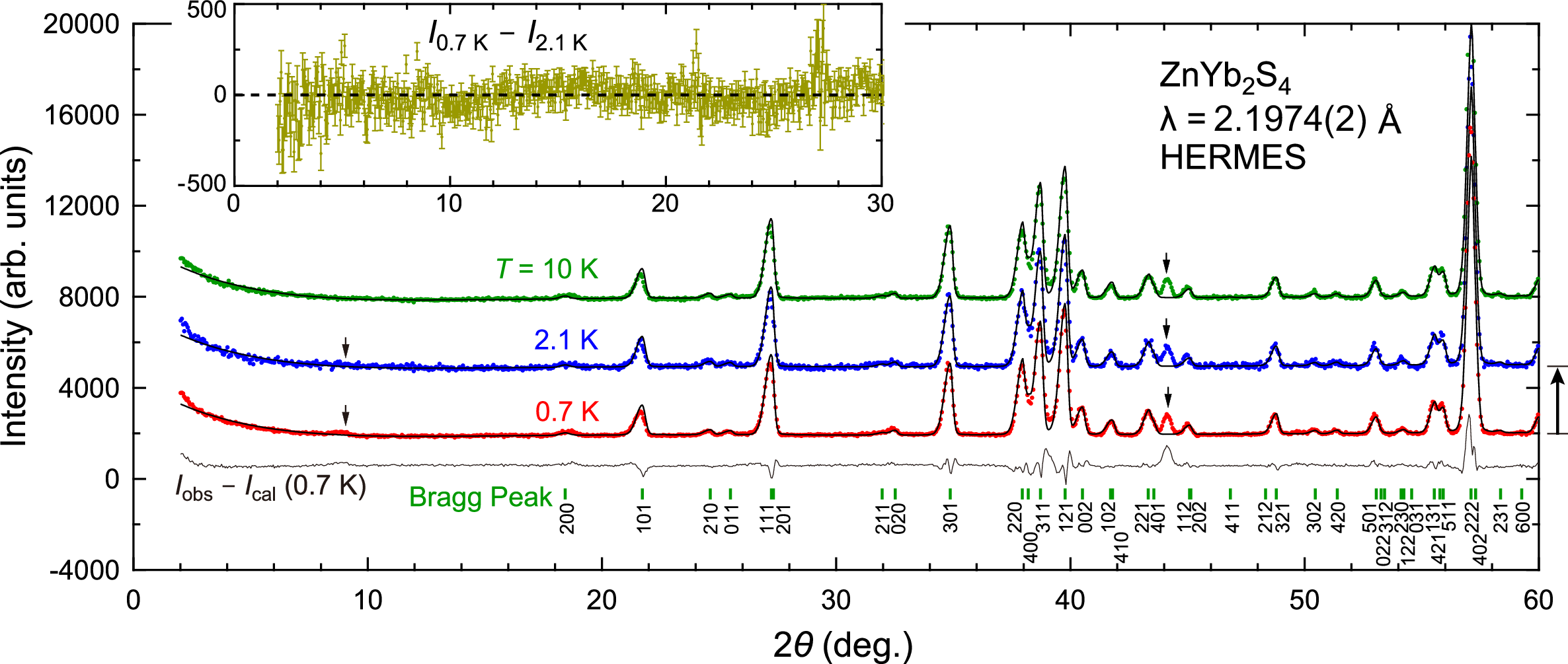}
    \caption{(Color online) Powder neutron diffraction patterns of ZnYb$_2$S$_4$ at $T$ = 0.7 (red), 2.1 (blue), and 10 K (green). 
    The intensities of the patterns at 2.1 and 10 K are normalized to the monitor counts at $T$ = 0.7 K, and the patterns at 2.1 and 10 K are vertically offset for clarity.
     The arrows denote impurity contributions at 8.9$^{\circ}$ and 44.1$^{\circ}$.
    The black lines represent the Rietveld refinement calculations, and the difference between the calculation and the data at 0.7 K is shown with the thin solid line below the measured patterns.
    The inset shows the difference in intensity between $T$~$=$~0.7 K and 2.1 K. 
    }
    \label{fig6}
\end{figure*}

\begin{table}[b]
  \centering
  \caption{Lattice constants, unit-cell volume, and the atomic coordinates of Yb(1), (0, 0, 0), and Yb(2), ($x_{\rm Yb(2)}$,~0.25,~$z_{\rm Yb(2)}$), of ZnYb$_2$S$_4$ determined by Rietveld analysis of powder neutron diffraction patterns measured at $T$ = 0.7, 2.1, and 10 K. $R_{\rm wp}$, $R_{\rm p}$, and $R_{\rm B}$ are the reliability factors for the Rietveld refinements. }
  \vspace{2mm}
\small
 \renewcommand{\arraystretch}{1.2}
  \begin{tabular}{llll}
    \hline
    $T$ (K) & \multicolumn{1}{c}{0.7} & \multicolumn{1}{c}{2.1} & \multicolumn{1}{c}{10.0} \\ \hline
    Monitor (cnts.) \hspace{1mm}      & \multicolumn{1}{c}{180000} & \multicolumn{1}{c}{30000} & \multicolumn{1}{c}{126000} \\
    $a$ (\AA)          & 13.2356(4) & 13.2364(5) & 13.2353(4) \\
    $b$ (\AA)          & \hspace{0.5mm} 7.7064(3)  & \hspace{0.5mm}  7.7065(3)  & \hspace{0.5mm}  7.7061(3)  \\
    $c$ (\AA)          & \hspace{0.5mm}  6.2606(2)  & \hspace{0.5mm}  6.2608(3)  & \hspace{0.5mm}  6.2605(2)  \\
    $V$ (\AA$^3$)      & \multicolumn{1}{c}{638.57(4)}  & \multicolumn{1}{c}{638.65(4)}  & \multicolumn{1}{c}{638.53(4)}  \\
    $x_{\rm Yb(2)}$ & 0.2335(2) & 0.2335(2)  & 0.2334(2) \\
    $z_{\rm Yb(2)}$ & 0.5034(9)  & 0.5047(10) & 0.5047(9)  \\
    $R_{\rm wp}$ (\%) & 12.9 & 15.3 & 13.0 \\
    $R_{\rm p}$ (\%)  & \hspace{0.5mm} 8.9  & 11.4 & \hspace{0.5mm}  9.0  \\
    $R_{\rm B}$ (\%)  & \hspace{0.5mm} 6.2  & \hspace{0.5mm} 7.0  & \hspace{0.5mm}  6.8  \\
    \hline
  \end{tabular}
  \label{t1}
\end{table}

  \subsection{Powder neutron diffraction}

To examine magnetic reflections below $T_{\rm m}$ and refine the structural parameters,
we carried out powder neutron diffraction experiments at $T$ $=$ 0.7, 2.1, and 10 K.
The resulting patterns are shown in Fig.~\ref{fig6}.
The intensities at 2.1 and 10 K were normalized to the monitor counts at $T$ = 0.7 K, as listed in Table~\ref{t1}, and their patterns in Fig.~\ref{fig6} are vertically offset for clarity.  
The black lines represent the Rietveld refinements, while the difference between the calculation and the 0.7 K data is shown as a solid line below the measured patterns.
Most of the peaks can be indexed using the orthorhombic crystal structure of ZnYb$_2$S$_4$ with the space group of $Pnma$, shown in Fig. \ref{fig1}(c).
The lattice constants, unit-cell volume, and atomic parameters for the Yb(2) site of ZnYb$_2$S$_4$, together with the reliability factors, are summarized in Table \ref{t1}.
The lattice constants at 0.7 K are determined as $a$~$=$~13.2356(4)~\AA, $b$~$=$~7.7064(3)~\AA, and $c$~$=$~6.2606(2)~\AA, which are nearly identical to those at 2.1 and 10.0 K. 
It is noted that the atomic coordinate for the Yb(2) site is equivalent within the error, indicating no distinct lattice modulation upon cooling below $T_{\rm m}$.
The black arrows at around $2\theta$ = 44.1$^{\circ}$ may indicate a temperature-independent extrinsic reflection from an impurity phase.
An additional reflection at 8.9$^{\circ}$ is seen at $T$ $=$ 0.7 K and 2.1 K, as shown with the arrows, although it likely arises from a magnetic impurity, since ZnYb$_2$S$_4$ orders at ${T}_{\rm m}$ = 1.4 K.

The upper inset of Fig. \ref{fig6} shows the intensity difference between $T$ = 0.7 K and 2.1 K, i.e., below and above $T_{\rm m}$, extracting the magnetic contribution associated with the magnetic order at $T_{\rm m}$.
No superlattice reflections were observed that would arise from antiferromagnetic modulation of the Yb moments, in line with the lack of a metamagnetic transition in the isothermal magnetization at $T$~$=$~0.28~K ($<$ $T_{\rm m}$) shown in Fig. \ref{fig5}.
On the other hand,
a subtle increase in intensity is observed around 27$^{\circ}$, near the 111 and 201 Bragg peaks,
while no other distinct magnetic reflections are extracted.
Although the error in the intensity data is considerably large, the measured intensity around 27$^{\circ}$ is still reasonable, 
provided that the Yb moments are uniformly aligned with a magnitude of 0.8(7) $\mu_{\rm B}$/Yb.
Here, it should be difficult to discuss the amplitude and alignment of the ordered Yb moments because of the inaccuracy of the estimated value; however, this observation is consistent with the ferromagnetic-like behavior inferred from our magnetization measurements (see Fig. \ref{fig5}).
To determine the magnetic structure, high-resolution diffraction measurements using a single crystal are highly desirable.

\subsection{Ground state of the Yb sawtooth chain}

As indicated by the experimental results such as the increase in the magnetization upon cooling below $T_{\rm m}$ (inset of Fig. \ref{fig2}), the reduced magnetic entropy at $T_{\rm m}$ (Fig.~\ref{fig3}), the magnetic field variation of  $T_{\rm m}$ (Fig.~\ref{fig4}), and the absence of superlattice reflections in the neutron diffraction measurements (Fig.~\ref{fig6}), the Yb moments are ferromagnetically aligned, although their magnitude is markedly reduced compared to the expected value ($\sim$1.33~$\mu_{\rm B}$/Yb) for the isolated doublet ground state, presumably due to magnetic frustration in the sawtooth Yb chain. 
The magnetic order with the reduced Yb ordered moments has been recently observed in the frustrated Yb zigzag chain system YbCuS$_2$ \cite{Onimaru25,Hori2023}.
As illustrated in the inset of Fig.~\ref{fig4}, it is expected in ZnYb$_2$S$_4$ that the reduced Yb moments (red arrows) are increased (blue arrows) due to release of the magnetic frustration in magnetic fields.
This uniform alignment of the reduced magnetic moments is distinct from the theoretically predicted ground states, such as the AFM order, the spin dimer state, and noncollinear order reminiscent of 1/2 magnetization plateau~\cite{Blundell03,Jiang15,Rausch25}.
This discrepancy may be attributed to additional exchange interactions between the Yb moments, e.g., inter-chain exchange interactions, which are neglected in the aforementioned theoretical models. 
Otherwise, theoretical analyses using Hubbard models for the quarter-filled insulating case indicate that the ground state should be ferromagnetic over a broad range of parameters, close to the paramagnetic regions with AFM correlations~\cite{Penc96}.
In ZnYb$_2$S$_4$, the AFM correlations in the paramagnetic state are suggested by the negative paramagnetic Curie temperature $\theta_{\rm p}$, while the ground state is characterized by ferromagnetically aligned Yb moments in the sawtooth chain.
Therefore, ZnYb$_2$S$_4$ presents the 4$f$-electron system exhibiting a characteristic ground state arising from magnetic frustration in a sawtooth chain.
Further microscopic measurements of spin correlations and investigations of anisotropic magnetic properties using a single-crystalline sample are indispensable.

 \section{Summary}

In conclusion, we have investigated the magnetic property of polycrystalline samples of ZnYb$_2$S$_4$, which contains sawtooth chains of Yb$^{3+}$ along the $b$-axis. 
The magnetic susceptibility, $M(T)/B$, follows the Curie--Weiss law for Yb$^{3+}$ in the temperature range of 70 $\le$ $T$ $\le$ 300 K, and the negative value of ${\it \theta}_{\rm p}$ indicates antiferromagnetic interactions between the Yb$^{3+}$ moments. 
The CEF ground state of Yb$^{3+}$ is a well-isolated Kramers doublet, giving rise to an effective spin-1/2 system.
The specific heat, $C$($T$), exhibits a sharp peak at $T_{\rm m}$~$=$~1.4~K, indicating a phase transition.
The magnetic entropy $S_{\rm m}$ at $T_{\rm m}$ is only 27\% of $R$ln2, which is smaller than the value expected for a Kramers doublet.
Therefore, a portion of $S_{\rm m}$ in the Kramers doublet is released even above $T_{\rm m}$ due to magnetic frustration in the sawtooth Yb chain. 
As the magnetic field increases, the peak in $C(T)$ broadens and shifts to higher temperatures.
The isothermal magnetization, $M$($B$), shows hysteresis for $T$ $<$ $T_{\rm m}$, with a spontaneous magnetization of 0.1 ${\it \mu}_{\rm B}$/Yb, and for $B$~$>$~0.2~T, $M(B)$ increases monotonously, reaching 1.1~${\it \mu}_{\rm B}$$/$Yb at 9 T. 
Powder neutron diffraction revealed no superlattice reflections associated with antiferromagnetic order, and instead, a subtle intensity increase was observed at Bragg positions, suggesting a ferromagnetic order of small Yb moments.
These observations indicate that, in the ground state of ZnYb$_2$S$_4$, the Yb moments reduced by magnetic frustration within the sawtooth Yb chain become ferromagnetically aligned, and are gradually recovered upon applying magnetic fields.
Thereby, the present study clarifies that ZnYb$_2$S$_4$ is the first 4$f$-electron system exhibiting characteristic properties arising from magnetic frustration in the sawtooth chain.
To elucidate the distinctive feature of the frustrated sawtooth Yb chain, it is indispensable to conduct experimental investigations of spin correlations and magnetic anisotropy using single crystals. 
\\

\section*{Acknowledgments}

The authors would like to thank K. Oda, R. Oishi, K. Umeo, and C. Hotta for the helpful discussion. 
The authors thank Y. Shibata for the electron-probe microanalysis carried out at N-BARD, Hiroshima University. 
The measurements with MPMS, PPMS, and the $^3$He heliox refrigerator were performed at N-BARD, Hiroshima University. 
We thank M. Ohkawara for their support for the neutron diffraction experiment at HERMES, JRR-3. 
The neutron diffraction experiments were carried out under the general user program managed by the Institute for Solid State Physics, University of Tokyo (Proposal No.~25588) and the Institute for Materials Research, Tohoku University (Proposal No. 202412-QBKNE-0019).
This work was financially supported by grants-in-aid from MEXT/JSPS of Japan [Grant Nos. JP15H05886, JP21K03440, JP22K03529, JP24H01673, and JP24K00574], and Grant-in-Aid for Transformative Research Areas (A) ``Asymmetric Quantum Matters'', JSPS KAKENHI Grant No. JP23H04870, and JST FOREST Program JPMJFR2233, and by The Kyoto Technoscience Center No. 8 in 2024, Japan.

\bibliography{./cite.bib}

\end{document}